\documentclass{PoS}
\usepackage{graphicx}
\usepackage{gensymb}
\usepackage{bm}
\usepackage{mathrsfs} 

\newcommand{\abs}[1]{\left| #1 \right|}
\newcommand{\pPb}{\textnormal{p--Pb}}
\newcommand{\PbPb}{\textnormal{Pb--Pb}}

\newcommand{\pT}{\ensuremath{p_\mathrm{T}}}

\title{Centrality dependence of charged jets in p--Pb collisions at $\sqrt{s_\mathrm{NN}} = 5.02$ TeV measured with the ALICE detector}

\ShortTitle{Centrality-dependent charged jets in p--Pb}

\author{\speaker{R\"udiger Haake} for the ALICE Collaboration\\
        Westf\"alische Wilhelms-Universit\"at M\"unster, Germany\\
        E-mail: \email{ruediger.haake@uni-muenster.de}}

\abstract{
Highly energetic jets are sensitive probes for the kinematics and the topology of nuclear collisions. Jets are collimated sprays of charged and neutral particles, which are produced in the fragmentation of hard scattered partons in an early stage of the collision. The measurement of jet spectra in p--Pb collisions provides an important way of quantifying the effects of cold nuclear matter in the initial state on jet production, fragmentation, and hadronization. Unlike in Pb--Pb collisions, strong hot nuclear matter effects -- e.g. from quark-gluon plasma formation -- are not expected to occur in p--Pb collisions. Hence, cold nuclear matter effects can be investigated in isolation.

The impact of cold nuclear matter effects on charged jet spectra is expected to depend on the event centrality. Higher event centralities are principally connected to a higher probability for an interaction of proton and lead-nucleus and therefore also for a possible nuclear modification. 

This article is the conference proceeding of a talk, in which centrality-dependent properties of charged jets in p--Pb measured by ALICE were shown for the first time. The focus is here on the fully corrected jet production cross sections and the nuclear modification factors. Additionally, the jet radial structure is explored by comparing jet spectra reconstructed with different resolution parameters.
}

\FullConference{53rd International Winter Meeting on Nuclear Physics,\\
		26-30 January 2015\\
		Bormio, Italy}

\begin{document}

\section{Introduction}

Jets can conceptually be described as the final state produced in a hard scattering. Therefore, jets are an excellent tool to access a very early stage of the collision. The jet constituents represent the final state remnants of the fragmented and hadronized partons that were scattered in the reaction. While all the detected particles have been created in a non-perturbative process (i.e. by hadronization), ideally, jets represent the kinematic properties of the originating partons. Thus, jets are mainly determined by perturbative processes due to the high momentum transfer and the cross sections can be calculated with pQCD. This conceptual definition is descriptive and very simple, the technical analysis of those objects is quite complicated though.\\

Proton--lead collisions are especially interesting to investigate how partonic interactions change when they take place in a nuclear environment. Besides, the measurements in p--Pb collisions also provide an important reference measurement for Pb-Pb collisions.

This article presents the centrality-dependent charged jet results in proton--lead collisions measured at 5.02 TeV with the ALICE experiment. Results and conclusions of this analysis are presented in Sec.~\ref{sec:Results}. Beforehand, a short description of the ALICE detector with regard to jet measurements is given in Sec.~\ref{sec:ALICEDetector} and important details of the analyses are shown in Sec.~\ref{sec:Techniques}. Additionally, the utilized centrality determination approach is presented very briefly in Sec.~\ref{sec:Centrality}.\\

\section{The ALICE detector}
\label{sec:ALICEDetector}
ALICE is the dedicated heavy-ion experiment at the LHC studying properties of the quark-gluon plasma and the QCD phase diagram in general. The detector is designed as a general-purpose heavy-ion detector \cite{ALICE2008} to measure and identify hadrons, leptons, and also photons down to very low transverse momenta.\\

One of ALICE's strengths is the excellent charged particle tracking capability. The tracking is performed in the central barrel, which consists of the Inner Tracking System (ITS) and the Time Projection Chamber (TPC). The ITS \cite{ALICE2010} is a cylindrical six-layered device consisting of three different semiconductor subdetectors: silicon pixel, drift, and strip detectors (SPD, SDD, and SSD). It directly surrounds the beam pipe. Around the ITS, the Time Projection Chamber is placed. The ALICE TPC \cite{ALICE2010b} is mainly filled with neon gas at atmospheric pressure, the gas detector has a radius of $250\mathrm{~cm}$ and a length of $500\mathrm{~cm}$. Combining the ITS and the TPC, a straight track can be reconstructed within a pseudorapidity interval of $\abs{\eta} < 0.9$. These tracks are the basic components of the studied charged jets.

For event triggering, the VZERO \cite{ALICE2010c} scintillation counters are utilized. The centrality estimation method makes use of the Zero-Degree Calorimeter (ZDC), a quartz fibers sampling calorimeter 116 m away from the interaction point.\\

\section{Jet reconstruction and correction techniques}
\label{sec:Techniques}

Jet reconstruction is usually a multi-step procedure. Apart from the conceptual definition of jets as hadronized partons scattered in a high energy collision, a technical jet concept is necessary. There is no unambiguous jet definition and no common way how to measure a jet. The currently most prevalently used jet algorithm at the LHC is the anti-$k_\mathrm{T}$ algorithm \cite{Cacciari2008} implemented in the FastJet \cite{Cacciari2006} package. For the presented analysis, the anti-$k_\mathrm{T}$ algorithm is applied to measure signal jets. Additionally, the $k_\mathrm{T}$ algorithm is utilized for the background correction technique.

The input passed to the jet finding algorithm is given by charged particles reconstructed with the TPC and ITS. The track cuts are chosen to obtain a uniform charged track distribution in the full $\eta-\phi$ plane.
Additionally, only tracks with $\abs{\eta} < 0.9$ and with $\pT > 150 \mbox{ MeV/}c$ are used in the jet finding procedure.

After constituent selection, the tracks are clustered by FastJet assuming massless jets. To avoid edge effects, only jets fully contained within the acceptance are used for further analysis.\\

Subsequently, two corrections are applied to the raw jets: background correction, which includes the subtraction of the mean event background density and the consideration of the background fluctuations within the event, and the correction for detector effects.

\subsection{Background correction}

In principle, every particle that does not originate from the hard parton scattering in question can be considered as background. Like the definition of a jet, also the definition of the background is ambiguous.
Several background correction techniques have been tested to find the most appropriate method.\\

The general ansatz is based on \cite{KTBackgroundCMS}. While the background density is calculated on an event-by-event basis, the background subtraction is performed depending on the area of the jet on a jet-by-jet basis.
To evaluate the average background transverse momentum density $\rho$, clusters found by the $k_\mathrm{T}$ algorithm are utilized. After excluding the two clusters with the highest transverse momenta, the median of the transverse momentum densities of the remaining $k_\mathrm{T}$ clusters is calculated. This average transverse momentum density is then scaled by a correction factor which takes into account how densely the acceptance is occupied by tracks. Thus, the background density is defined by
\begin{equation}
\rho = \mathrm{median} \left\{ \frac{p^\mathrm{clus}_{\mathrm{T}, i} }{A^\mathrm{clus}_i} \right\} \cdot C.
\end{equation}
Here, $p^\mathrm{clus}_{\mathrm{T}, i}$ and $A^\mathrm{clus}_i$ represent the cluster's transverse momentum and area, respectively. $C$ is the occupancy correction factor \cite{KTBackgroundCMS}. It is defined by the ratio of area that contains tracks over the full acceptance area.

To subtract the background momentum on a jet-by-jet basis, the jet area must be known. The correction is then applied by
\begin{equation}
p^\mathrm{corr}_\mathrm{T,\;jet} = p^\mathrm{orig}_\mathrm{T,\;jet} - \rho A.
\end{equation}\\

Region-to-region fluctuations of the background are taken into account by probing the transverse momentum in randomly distributed cones and comparing it to the average background.
Quantitatively, this distribution is given by
\begin{equation}
\delta \pT = \sum_\mathrm{i}{p_\mathrm{T,\,i}-\rho A}, ~~~ A = \pi R^2.
\end{equation}
In the formula, $R$ represents the radius of the cone.\\
While the background subtraction can be applied separately for every jet, the background fluctuations can only be taken into account on a probabilistic basis in an unfolding procedure.

\subsection{Detector effects}

Like the background fluctuations, detector effects -- e.g. from the limited single-particle tracking efficiency -- are considered in an SVD unfolding procedure \cite{SVDAlgorithm}.\\
The detector response matrix utilized in the unfolding is created with a full detector simulation using PYTHIA6 to generate jets and GEANT3 for the particle transport through the detector. Several cross checks including toy model analyses have been performed to guarantee an optimum configuration of the unfolding approach and to properly estimate the systematic uncertainty.\\

\section{Centrality determination in p--Pb collisions with ALICE}
\label{sec:Centrality}

Event centrality is a measure for the overlap of two nuclei in a collision, describing the collision geometry.
In Pb--Pb, the multiplicity at midrapidity is tightly connected to the centrality. Therefore, the events can be directly classified by dividing the full dataset according to the measured charged particle multiplicity distribution.\\
For the evaluation of the number of binary collisions $N_\mathrm{coll}$ and participants $N_\mathrm{part}$, a Glauber Monte Carlo (MC) simulation is performed to simulate $N_\mathrm{part}$ depending on the impact parameter $b$. Since $b$ is not directly measurable, the $N_\mathrm{part}$ distribution is connected to a multiplicity distribution by a Negative Binomial Distribution (NBD) fit which applies a simple model for particle production \cite{CentralityPbPb}.\\

Unfortunately, this approach is not directly applicable in $\pPb$ collisions. The main problem is that the correlation of multiplicity and centrality is much worse than in $\PbPb$. This results e.g. in high-multiplicity upward fluctuations of peripheral events that are then characterized as central events.\\

In \cite{CentralityPaper}, a hybrid centrality estimation approach is introduced that makes use of the ZDC. Instead of connecting a multiplicity distribution at midrapidity to centrality, the deposited energy in the ZDC is used. However, this energy distribution cannot be connected to $N_\mathrm{part}$ by applying a simple NBD fit. More complex models on slow nucleon emission would be necessary to estimate the particle abundance in the ZDC. This would introduce a larger systematic uncertainty and, therefore, a different ansatz was chosen.
The number of binary collisions $N_\mathrm{coll}$ is calculated using different experimental observables and certain assumptions on their centrality dependence.
For jet analyses, two estimators have been considered, namely

\begin{itemize}
  \item $N_\mathrm{coll}^\mathrm{mult}$: This estimator assumes that the charged particle multiplicity at midrapidity is proportional to $N_\mathrm{part}$,
  \item $N_\mathrm{coll}^\mathrm{Pb-side}$: The charged particle multiplicity on the Pb-going side is assumed to be proportional to the number of participating target nucleons $N_\mathrm{part} -1 = N_\mathrm{coll}$.
\end{itemize}
In the present analysis, both estimators are treated as valid choices. However, $N_\mathrm{coll}^\mathrm{Pb-side}$ is conceptually probably the better choice: While $N_\mathrm{coll}^\mathrm{mult}$ is calculated using mean $\mathrm{d}N/\mathrm{d}\eta$ distributions obtained by TPC and ITS, $N_\mathrm{coll}^\mathrm{Pb-side}$ is estimated using data from the VZERO detector which is probably less affected by midrapidity fluctuations.
A more detailed description of the centrality estimators can be found in \cite{CentralityPaper}.\\

\section{Results}
\label{sec:Results}

\begin{figure}[!htp]
\centering
\includegraphics[width=1.0\textwidth]{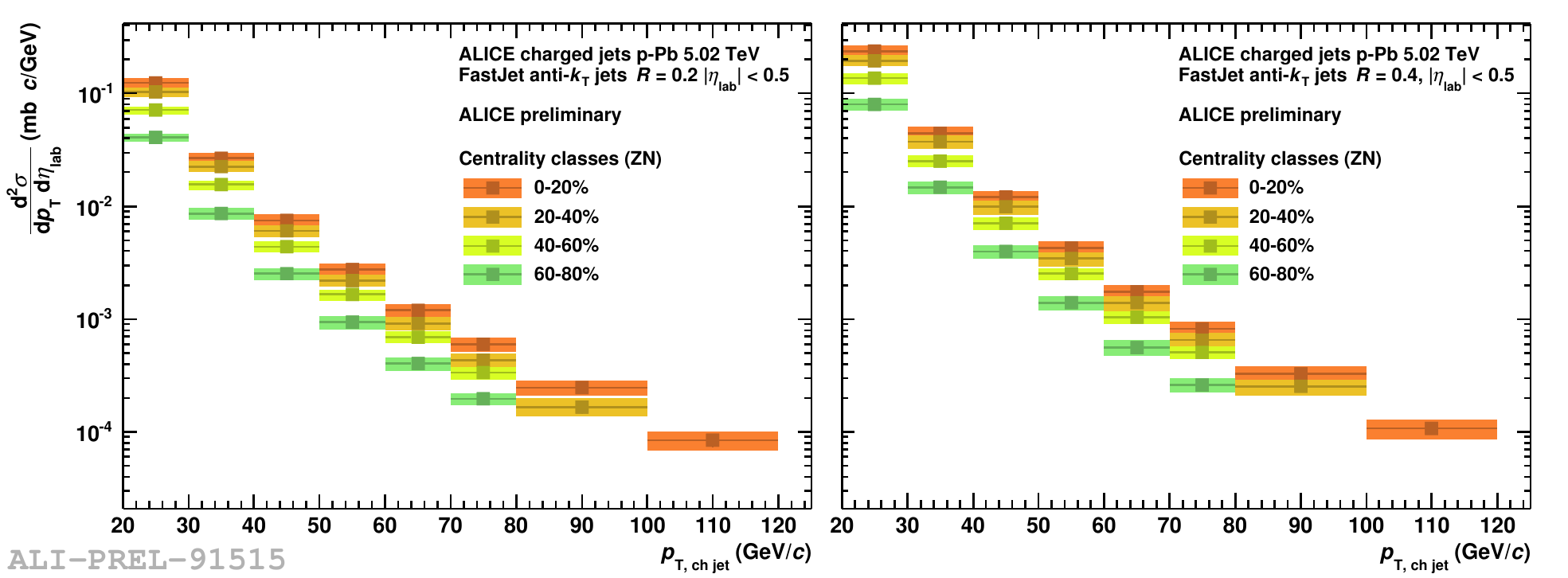}
\caption{Charged jet production cross sections in p--Pb collisions for selected centrality intervals for $R=0.2$ and $R=0.4$. Note that the 3.5\% normalization uncertainty of the total cross section measurement is not shown.}
\label{fig:Spectra_ZNA_AllCentralities}
\end{figure}

In the following, the results on centrality-dependent jet production are presented. Preliminary results for minimum bias collisions have already been published \cite{PrelimMB} and will not be shown.
Figure \ref{fig:Spectra_ZNA_AllCentralities} shows the fully corrected jet production (visible) cross sections. To apply the correct normalization, the total visible cross section in $\pPb$ is needed. It was measured to be $\sigma_\mathrm{V0} = 2.09 \pm 0.07$ b \cite{CrossSectionPaper}. The term visible cross section refers to the cross section measured by the VZERO detector without any correction for the trigger efficiency. Both resolution parameters $R=0.2$ and $R=0.4$ are presented. For all shown observables, statistical uncertainties are depicted as errors bars and systematic uncertainties are represented by filled boxes around the data points.\\

In $\pPb$ collisions, the (centrality-dependent) nuclear modification factor $Q_\mathrm{pPb}$ is a measure for the effect of the nuclear environment of the collision on jet production. The factor is defined by
\begin{equation}
Q_\mathrm{pPb} =
\frac
{\left.\frac{\mathrm{d}N}{\mathrm{d}p_\mathrm{T}\mathrm{d}\eta} \right| _\mathrm{pPb, cent}}
{\left<T_\mathrm{pPb}^\mathrm{cent}\right>   \cdot \left.\frac{\mathrm{d}\sigma}{\mathrm{d}p_\mathrm{T}\mathrm{d}\eta} \right| _\mathrm{pp}}.
\end{equation}
$T_\mathrm{pPb}^\mathrm{cent}$ is the nuclear overlap function \cite{CentralityPaper}, which connects the $\pPb$ collision system to a reference measurement in pp. By construction, the nuclear modification factor is one if no nuclear effects are present.
Since there is currently no pp collision data available for $\sqrt{s_\mathrm{NN}} = 5.02$ TeV, the pp reference was constructed using (scaled) charged jets measured at 7 TeV \cite{PrelimMB}.\\

\begin{figure}[!htp]
\centering
\includegraphics[width=1.0\textwidth]{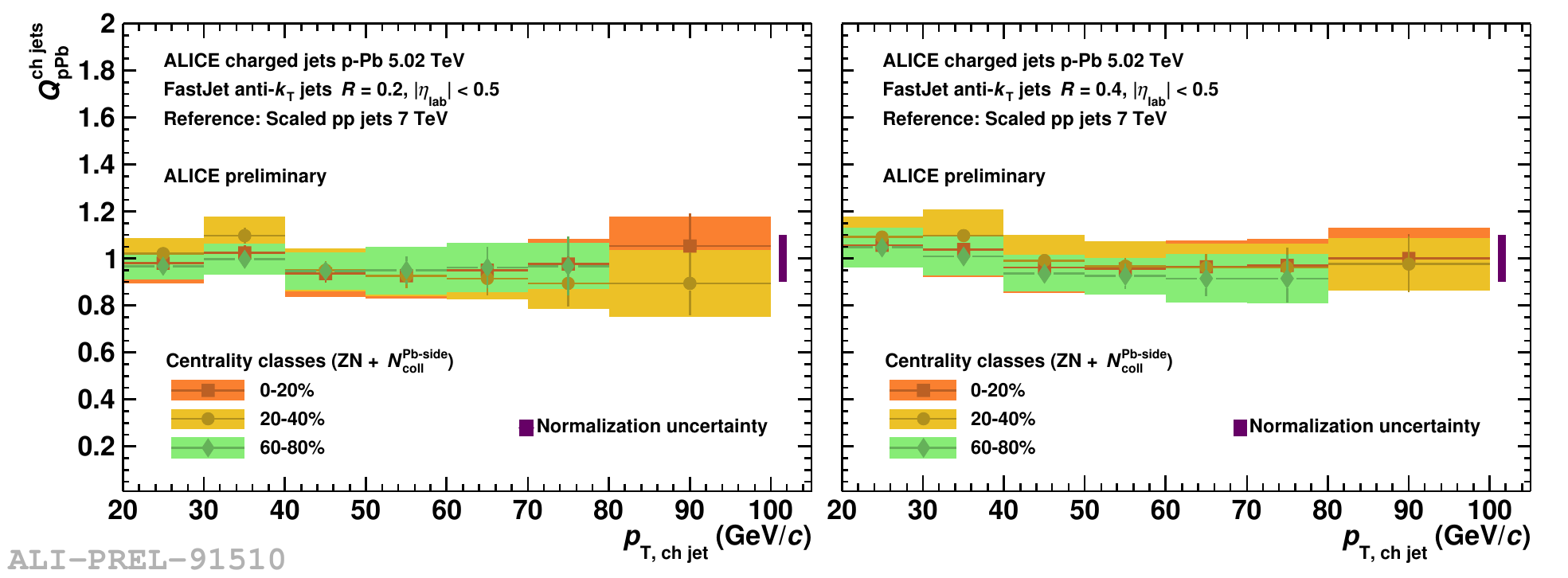}
\caption{Charged jet nuclear modification factors $Q_\mathrm{pPb}$ for selected centrality intervals for $R=0.2$ and $R=0.4$ using the $N_\mathrm{coll}^\mathrm{Pb-side}$ estimator.}
\label{fig:QpPb_ZNA_PbSide_AllCentralities}
\includegraphics[width=1.0\textwidth]{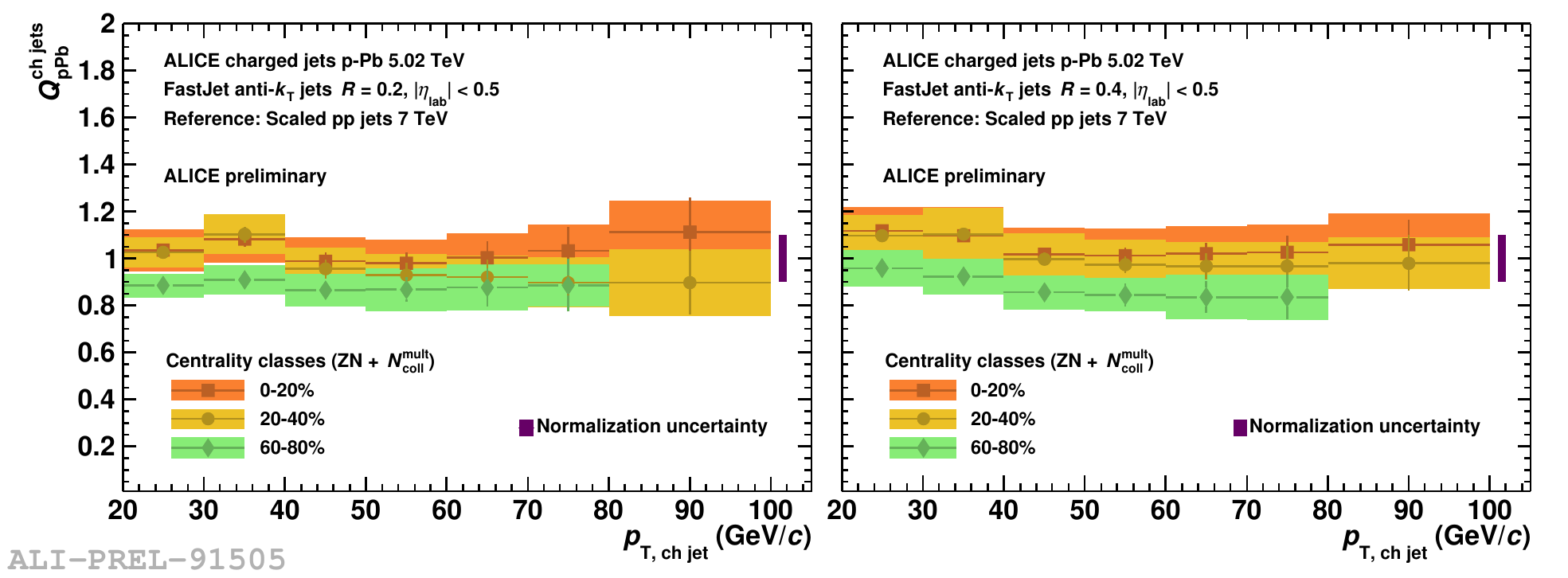}
\caption{Charged jet nuclear modification factors $Q_\mathrm{pPb}$ for selected centrality intervals for $R=0.2$ and $R=0.4$ using the $N_\mathrm{coll}^\mathrm{mult}$ estimator.}
\label{fig:QpPb_ZNA_AllCentralities}
\end{figure}

In Figs.~\ref{fig:QpPb_ZNA_PbSide_AllCentralities} and \ref{fig:QpPb_ZNA_AllCentralities}, $Q_\mathrm{pPb}$ is shown for the $N_\mathrm{coll}^\mathrm{Pb-side}$ and $N_\mathrm{coll}^\mathrm{mult}$ estimator, respectively.\\
Within the uncertainties, the nuclear modification factor shows no dependence on centrality for both $N_\mathrm{coll}$ estimates and the different centrality classes. For the $N_\mathrm{coll}^\mathrm{mult}$, a slightly stronger spread of the different distributions can be observed. There is no indication of strong nuclear effects like jet suppression or enhancement for all centrality classes.
This also holds for the centrality classes that have been left out in the figures for the reason of better visibility.\\

To some extent, the difference between the $Q_\mathrm{pPb}$'s from both estimates can be seen as a measure for the systematic uncertainty of the hybrid centrality approach: Both $N_\mathrm{coll}$ values are valid choices of the centrality for the jet measurement scenario.\\
In order to measure the radial jet structure and jet collimation, the jet cross section ratio has been measured for several centrality classes. It is defined by
\begin{equation}
\label{eq:JetShape}
\mathscr{R}(0.2,\,0.4) = \frac{\mathrm{d}\sigma_{\mathrm{pPb,\,} R=0.2} / \mathrm{d}p_\mathrm{T}}{\mathrm{d}\sigma_{\mathrm{pPb,\,} R=0.4} / \mathrm{d}p_\mathrm{T}}.
\end{equation}
Note that this observable is just a very rough measure of the radial jet structure.\\

\begin{figure}
\centering
\includegraphics[width=0.94\textwidth]{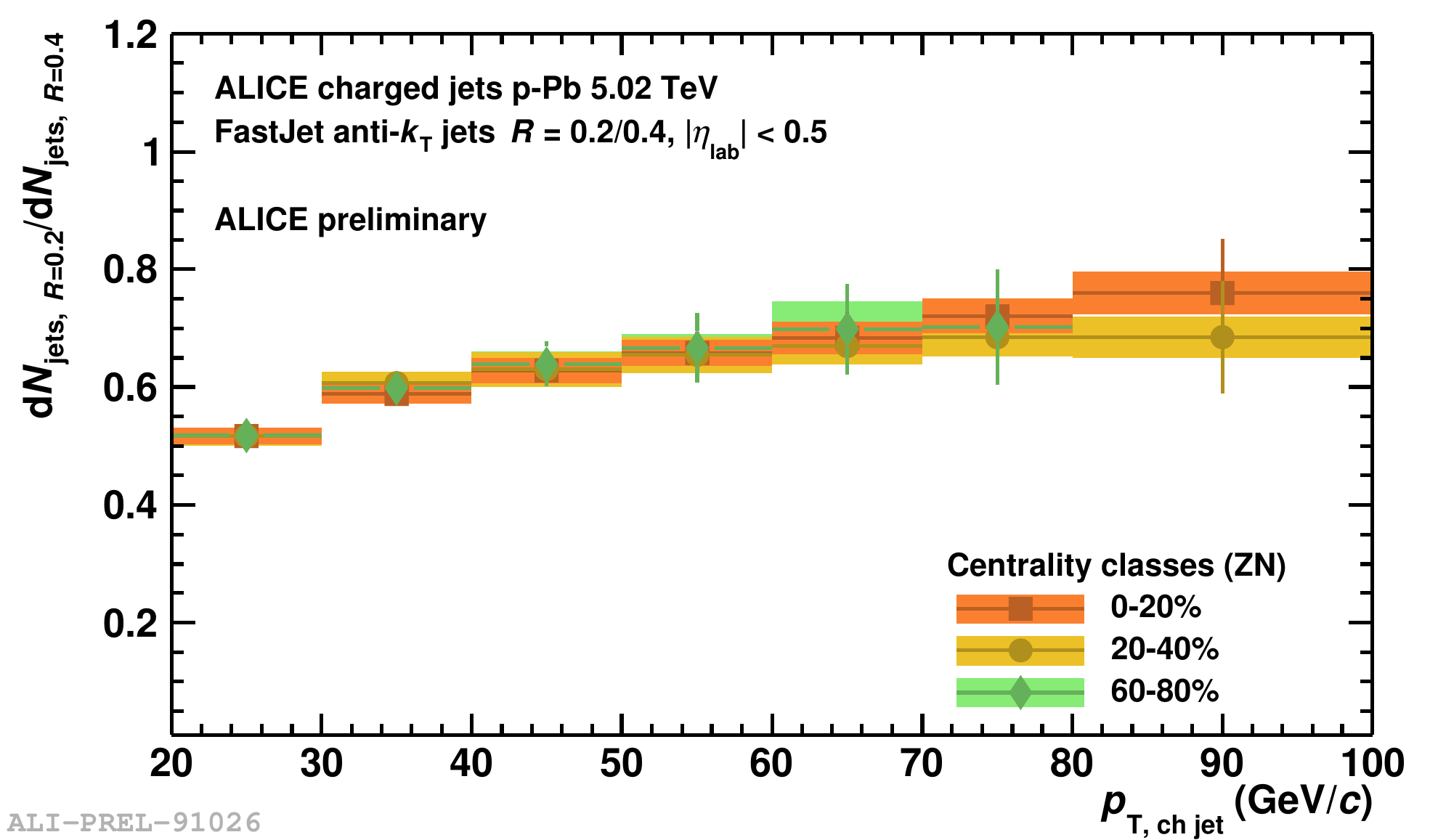}
\caption{Charged jet cross section ratio in p--Pb collisions for selected centrality intervals for $R=0.2$ and $R=0.4$.}
\label{fig:Shape_ZNA_AllCentralities}
\end{figure}

In Fig.~\ref{fig:Shape_ZNA_AllCentralities}, the ratio is shown for different centrality classes. Within the uncertainties, no significant centrality dependence is observed for all classes, including those classes not explicitly shown.\\
This result is fully compatible with the expectations: Even in $\PbPb$ collisions, no significant jet structure modification was observed for jets \cite{ALICE2014a}.\\

\section{Summary}

Centrality-dependent charged jets in $\pPb$ collisions were measured by ALICE using the novel hybrid centrality approach. Jet production cross sections were presented for the resolution parameters $R=0.2$ and $R=0.4$ for several centrality classes. Using two different estimates for the number of binary collisions $N_\mathrm{coll}$, the nuclear modification factor $Q_\mathrm{pPb}$ was shown to be compatible only with small or even no centrality dependence.\\
Therefore, there are also no hints for strong nuclear effects on jet production in $\pPb$. Additionally, the jet cross section ratio as a simple measure for the radial jet structure shows no significant modification with respect to centrality.

\begin{thebibliography}{99}

\bibitem{ALICE2008}
  ALICE Collaboration:
  \emph{The ALICE experiment at the CERN LHC},
  \emph{JINST} {\bf{3}} (2008) 8002.

\bibitem{ALICE2010}
  ALICE Collaboration:
  \emph{Alignment of the ALICE Inner Tracking System with cosmic-ray tracks},
  \emph{JINST} {\bf{5}} (2010) 3003,
  {\tt [physics.Ins-det/1001.0502]}.

\bibitem{ALICE2010b}
  ALICE Collaboration:
  \emph{The ALICE TPC, a large 3-dimensional tracking device with fast readout for ultra-high multiplicity events},
  \emph{Nucl. Instrum. Meth. A} {\bf{622}} (2010) 316-367,
  {\tt [physics.Ins-det/1001.1950]}.

\bibitem{ALICE2010c}
  ALICE Collaboration:
  \emph{Charged-Particle Multiplicity Density at Midrapidity in Central Pb--Pb Collisions at $\sqrt{s_\mathrm{NN}}=2.76$ TeV},
  \emph{Phys. Rev. Lett.} {\bf{105}} (2010) 252301,
  {\tt [nucl-ezx/1011.3916]}.

\bibitem{Cacciari2008}
  M. Cacciari / G.P. Salam / G. Soyez:
  \emph{The anti-$k_T$ jet clustering algorithm},
 	\emph{JHEP} {\bf 0804} (2008) 063,
  {\tt [hep-ph/0802.1189]}.

\bibitem{Cacciari2006}
  M. Cacciari / G.P. Salam:
  \emph{Dispelling the $N^3$ myth for the $k_t$ jet-finder},
 	\emph{Phys. Lett. B} {\bf 641} 57-61,
  {\tt [hep-ph/0512210]}.

\bibitem{KTBackgroundCMS}
  The CMS Collaboration:
  \emph{Measurement of the underlying event activity in pp collisions at $\sqrt{s} = 0.9$ and 7 TeV with the novel jet-area/median approach},
  \emph{JHEP} {\bf 08} (2012) 130,
  {\tt [hep-ex/1207.2392]}.

\bibitem{SVDAlgorithm}
  A. Hoecker / V. Kartvelishvili:
  \emph{SVD Approach to Data Unfolding},
  \emph{Nucl. Instrum. Meth.} {\bf A372} (1996) 469-481,
  {\tt [hep-ph/9509307]}.

\bibitem{CentralityPbPb}
  ALICE Collaboration:
  \emph{Centrality determination of Pb--Pb collisions at $\sqrt{s_\mathrm{NN}}$ = 2.76 TeV with ALICE},
  \emph{Phys. Rev. C} {\bf 88} (2013) 44909,
  {\tt [nucl-ex/1301.4361]}.

\bibitem{CentralityPaper}
  ALICE Collaboration:
  \emph{Centrality dependence of particle production in p-Pb collisions at $\sqrt{s_\mathrm{NN}}$ = 5.02 TeV},
  \emph{submitted to Phys. Rev. C} (2014),
  {\tt [nucl-ex/1412.6828]}.

\bibitem{CrossSectionPaper}
  ALICE Collaboration:
  \emph{Measurement of visible cross sections in proton-lead collisions at $\sqrt{s_\mathrm{NN}} = 5.02$ TeV in van der Meer scans with the ALICE detector},
  \emph{JINST} {\bf 9} (2014) 1100,
  {\tt [nucl-ex/1405.1849]}.

\bibitem{PrelimMB}
  R. Haake for the ALICE Collaboration:
  \emph{Charged Jets in Minimum Bias p--Pb Collisions at $\sqrt{s_\mathrm{NN}} = $ 5.02 TeV with ALICE},
  \emph{PoS EPS-HEP2013} (2013) 176,
  {\tt [nucl-ex/1310.3612]}.

\bibitem{ALICE2014a}
  ALICE Collaboration:
  \emph{Measurement of charged jet suppression in Pb--Pb collisions at $\sqrt{s_\mathrm{NN}}=$ 2.76 TeV},
  \emph{JHEP} {\bf 30} (2014) 013,
  {\tt [nucl-ex/1311.0633]}.

\end{thebibliography}
\end{document}